
\hsize=6.0in \hoffset=0.25in \tolerance=1000\hfuzz=2pt


\def\monthintext{\ifcase\month\or January\or February\or
   March\or April\or May\or June\or July\or August\or
   September\or October\or November\or December\fi}




\def\preprint#1{\baselineskip=18pt\pageno=0
   \begingroup
   \vsize=8.0in\voffset=0.5in\nopagenumbers\parindent=0pt\baselineskip=14.4pt
   \rightline{#1}}

\def\date#1{\rightline{#1}}

\def\title#1#2{
   \vskip 0.7in plus 0.2in
   \centerline{\titlefont #1}
   \vskip 0.2in
   \centerline{\titlefont #2}
   \vskip 0.8in plus 0.1in}

\def\author#1#2#3{\centerline{{\bf #1}\ulabelfoot{#2}{#3}}\smallskip}
\def\address#1{\centerline{#1}}
\def\abstract{\bigskip\bigskip\bigskip\medskip
	\centerline{\bf Abstract}
	\smallskip}
\def\finishtitlepage{\vskip 0.8in plus 0.3in
   \supereject\endgroup
   \baselineskip=18pt}

\def\nolabels{\def\eqnlabel##1{}\def\eqlabel##1{}\def\figlabel##1{}%
	\def\reflabel##1{}}
\def\writelabels{\def\eqnlabel##1{%
	{\escapechar=` \hfill\rlap{\hskip.11in\string##1}}}%
	\def\eqlabel##1{{\escapechar=` \rlap{\hskip.11in\string##1}}}%
	\def\figlabel##1{\noexpand\llap{\string\string\string##1\hskip.66in}}%
	\def\reflabel##1{\noexpand\llap{\string\string\string##1\hskip.37in}}}


\global\newcount\secno \global\secno=0
\global\newcount\meqno \global\meqno=1

\def\newsec#1{\global\advance\secno by1
\xdef\secsym{\the\secno.}\global\meqno=1
	\bigbreak\medskip
	\noindent{\bf\the\secno. #1}\par\nobreak\medskip\nobreak\noindent}
\xdef\secsym{}

\def\appendix#1#2{\global\meqno=1\xdef\secsym{\hbox{#1.}}\bigbreak\bigskip
\noindent{\bf Appendix #1. #2}\par\nobreak\medskip\nobreak}

\def\undersection#1{\par
   \ifnum\the\lastpenalty=30000\else \penalty-100\medskip \fi
   \noindent\undertext{#1}\enspace \vadjust{\penalty5000}}
\def\undertext#1{\vtop{\hbox{#1}\kern 1pt \hrule}}
\def\subsection#1{\undersection{#1} \medskip}
%

%
\def\ack{\ifnum\the\lastpenalty=30000\else \penalty-100\smallskip \fi
   \noindent\undertext{Acknowledgements:}\enspace \vadjust{\penalty5000}}
%

\def\eqnn#1{\xdef #1{(\secsym\the\meqno)}%
	\global\advance\meqno by1\eqnlabel#1}
\def\eqna#1{\xdef #1##1{\hbox{$(\secsym\the\meqno##1)$}}%
	\global\advance\meqno by1\eqnlabel{#1$\{\}$}}
\def\eqn#1#2{\xdef #1{(\secsym\the\meqno)}\global\advance\meqno by1%
	$$#2\eqno#1\eqlabel#1$$}


\def\ulabelfoot#1#2{{\tenpoint\baselineskip=12pt plus
0.2pt\footnote{#1}{#2}}}        
\global\newcount\ftno \global\ftno=1
\def\foot#1{{\tenpoint\baselineskip=12pt plus
0.2pt\footnote{$^{\the\ftno}$}{#1}}%
	\global\advance\ftno by1}


\global\newcount\refno \global\refno=1
\newwrite\rfile

\def\ref{[\the\refno]\nref}
\def\nref#1{\xdef#1{[\the\refno]}\ifnum\refno=1\immediate
	\openout\rfile=refs.tmp\fi\global\advance\refno by1\chardef\wfile=\rfile
	\immediate\write\rfile{\noexpand\item{#1\ }\reflabel{#1}\pctsign}\findarg}
\def\findarg#1#{\begingroup\obeylines\newlinechar=`\^^M\passarg}
	{\obeylines\gdef\passarg#1{\writeline\relax #1^^M\hbox{}^^M}%
	\gdef\writeline#1^^M{\expandafter\toks0\expandafter{\striprelax #1}%
	\edef\next{\the\toks0}\ifx\next\null\let\next=\endgroup\else\ifx\next\empty%
	\else\immediate\write\wfile{\the\toks0}\fi\let\next=\writeline\fi\next\relax}}
	{\catcode`\%=12\xdef\pctsign{
\def\striprelax#1{}

\def\semi{;\hfil\break}
\def\addref#1{\immediate\write\rfile{\noexpand\item{}#1}} 

\def\listrefs{\immediate\closeout\rfile
  \bigskip \centerline{{\bf References}}\medskip{\frenchspacing%
   \catcode`\@=11\escapechar=` %
   \input refs.tmp \vfill\eject}\nonfrenchspacing}

\def\startrefs#1{\immediate\openout\rfile=refs.tmp\refno=#1}


\global\newcount\figno \global\figno=1
\newwrite\ffile
\def\fig{\the\figno\nfig}
\def\nfig#1{\xdef#1{\the\figno}\ifnum\figno=1\immediate
	\openout\ffile=figs.tmp\fi\global\advance\figno by1\chardef\wfile=\ffile
	\immediate\write\ffile{\medskip\noexpand\item{Fig.\ #1:\ }%
	\figlabel{#1}\pctsign}\findarg}

\def\listfigs{\vfill\eject\immediate\closeout\ffile{\parindent48pt
	\baselineskip16.8pt\centerline{{\bf Figure Captions}}\medskip
	\escapechar=` \input figs.tmp\vfill\eject}}



\def\centerps#1{ \centerline{}}

\def\figcaption#1#2{\tenpoint\baselineskip12pt \noindent
{\bf Fig.~{#1}.}{\rm {\ #2} } }


\def\topfigure#1#2#3{\topinsert {\centerps{#1}}
\smallskip \figcaption {#2} {#3} \endinsert}


\def\tenpoint{\def\rm{\fam0\tenrm}
\textfont0=\tenrm\scriptfont0=\sevenrm\scriptscriptfont0=\fiverm
\textfont1=\teni\scriptfont1=\seveni\scriptscriptfont1=\fivei
\textfont2=\tensy\scriptfont2=\sevensy\scriptscriptfont2=\fivesy
\textfont\itfam=\tenit\def\it{\fam\itfam\tenit}%
\textfont\bffam=\tenbf\def\bf{\fam\bffam\tenbf}\rm}
%


\font\magtenrm=cmr10 scaled \magstep1
\font\magsevenrm=cmr7 scaled \magstep1
\font\magfiverm=cmr5 scaled \magstep1

\font\magtenbf=cmbx10 scaled \magstep1
\font\magsevenbf=cmbx7 scaled \magstep1
\font\magfivebf=cmbx5 scaled \magstep1

\font\magteni=cmmi10 scaled \magstep1
\font\magseveni=cmmi7 scaled \magstep1
\font\magfivei=cmmi5 scaled \magstep1

\font\magtensy=cmsy10 scaled \magstep1
\font\magsevensy=cmsy7 scaled \magstep1
\font\magfivesy=cmsy5 scaled \magstep1

\font\magtenex=cmex10 scaled \magstep1
\font\magtentt=cmtt10 scaled \magstep1
\font\magtenit=cmti10 scaled \magstep1
\font\magtensl=cmsl10 scaled \magstep1

\def\twelvepoint{\def\rm{\fam0\magtenrm}
	\textfont0=\magtenrm \scriptfont0=\magsevenrm \scriptscriptfont0=\magfiverm
	\textfont1=\magteni  \scriptfont1=\magseveni  \scriptscriptfont1=\magfivei
	\textfont2=\magtensy \scriptfont2=\magsevensy \scriptscriptfont2=\magfivesy
	\textfont\itfam=\magtenit \def\it{\fam\itfam\magtenit}
	\textfont\ttfam=\magtentt \def\tt{\fam\ttfam\magtentt}
	\textfont\bffam=\magtenbf \def\bf{\fam\bffam\magtenbf}
	\textfont\slfam=\magtensl \def\sl{\fam\slfam\magtensl} \rm
   \hfuzz=1pt\vfuzz=1pt
   \setbox\strutbox=\hbox{\vrule height 10.2pt depth 4.2pt width 0pt}
   \parindent=24pt\parskip=0pt plus 1.2pt
   \topskip=12pt\maxdepth=4.8pt\jot=3.6pt
	\normalbaselineskip=14.4pt\normallineskip=1.2pt
   \normallineskiplimit=0pt\normalbaselines
	\abovedisplayskip=13pt plus 3.6pt minus 5.8pt
   \belowdisplayskip=13pt plus 3.6pt minus 5.8pt
   \abovedisplayshortskip=-1.4pt plus 3.6pt
   \belowdisplayshortskip=13pt plus 3.6pt minus 3.6pt
   \topskip=12pt \splittopskip=12pt
   \scriptspace=0.6pt\nulldelimiterspace=1.44pt\delimitershortfall=6pt
   \thinmuskip=3.6mu\medmuskip=3.6mu plus 1.2mu minus 1.2mu
   \thickmuskip=4mu plus 2mu minus 1mu
   \smallskipamount=3.6pt plus 1.2pt minus 1.2pt
   \medskipamount=7.2pt plus 2.4pt minus 2.4pt
   \bigskipamount=14.4pt plus 4.8pt minus 4.8pt}

\twelvepoint



\font\titlerm=cmr10 scaled \magstep3
\font\titlerms=cmr10 scaled \magstep1

\font\titlei=cmmi10 scaled \magstep3  
\font\titleis=cmmi10	scaled \magstep1

\font\titlesy=cmsy10 scaled \magstep3 	
\font\titlesys=cmsy10 scaled \magstep1

\font\titleit=cmti10 scaled \magstep3	

\skewchar\titlei='177 \skewchar\titleis='177 
\skewchar\titlesy='60 \skewchar\titlesys='60 

\def\titlefont{\def\rm{\fam0\titlerm}
   \textfont0=\titlerm \scriptfont0=\titlerms 
   \textfont1=\titlei  \scriptfont1=\titleis  
   \textfont2=\titlesy \scriptfont2=\titlesys 
   \textfont\itfam=\titleit \def\it{\fam\itfam\titleit} \rm}


\def\noblackbox{\overfullrule=0pt}
\def\inv{^{\raise.18ex\hbox{${\scriptscriptstyle -}$}\kern-.06em 1}}
\def\dup{^{\vphantom{1}}}
\def\Dsl{\,\raise.18ex\hbox{/}\mkern-16.2mu D} 
\def\dsl{\raise.18ex\hbox{/}\kern-.68em\partial}
\def\slash#1{\raise.18ex\hbox{/}\kern-.68em #1}
\def\boxeqn#1{\vcenter{\vbox{\hrule\hbox{\vrule\kern3.6pt\vbox{\kern3.6pt
	\hbox{${\displaystyle #1}$}\kern3.6pt}\kern3.6pt\vrule}\hrule}}}
\def\mbox#1#2{\vcenter{\hrule \hbox{\vrule height#2.4in
	\kern#1.2in \vrule} \hrule}}  
\def\half{{\textstyle{1\over2}}}
\def\ha{{1\over2}}
\def\e#1{{\rm e}^{\textstyle#1}}
\def\C#1{\hbox{{$\cal #1$}}}    
\def\O{\hbox{{$\cal O$}}}	
\def\({\left(} \def\){\right)}  
\def\[{\left[} \def\]{\right]}  
\def\l|{\left|} \def\r|{\right|}
\def\vev#1{\langle #1 \rangle}
\def\<{\langle}
\def\>{\rangle}
\def\psibar{\overline\psi}
\def\lform{\hbox{$\sqcup$}\llap{\hbox{$\sqcap$}}}
\def\grad#1{\,\nabla\!_{{#1}}\,}
\def\gradgrad#1#2{\,\nabla\!_{{#1}}\nabla\!_{{#2}}\,}
\def\darr#1{\raise1.8ex\hbox{$\leftrightarrow$}\mkern-19.8mu #1}
\def\roughly#1{\raise.3ex\hbox{$#1$\kern-.75em\lower1ex\hbox{$\sim$}}}
\def\en{\eqalign}
\def\del{\partial}
\def\Tr{ {\rm Tr} }

\hyphenation{di-men-sion di-men-sion-al di-men-sion-al-ly
             di-men-sion-al-i-ty}
\nolabels

\def\xplus{ x^{+} }
\def\xminus{ x^{-} }
\def\xperp{ {{\vec x}_{\perp}} }
\def\bfalpha{ {\vec \alpha}}
\def\bfbeta{ {\vec \beta}}
\def\tGamma{ {\tilde \Gamma} }
\def\tLambda{ {\tilde \Lambda} }
\def\cJ{ {\cal J} }
\def\cE{ {\cal E} }
\def\cH{ {\cal H} }
\def\cS{ {\cal S} }
\def\bJ{ {\bar J} }
\def\bcJ{ {\bar {\cal J} } }
\def\tJ{ {\tilde J} }
\def\bK{ {\bar K} }
\def\bL{ {\bar L} }
\def\bF{ {\bar F} }
\def\bz{ {\bar z} }
\def\bw{ {\bar w} }
\def\bn{ {\bar n} }
\def\baell{ {\bar \ell} }
\def\bphi{ {\bar \phi} }
\def\cO{ {\cal O} }
\def\state#1{ {| #1\rangle} }
\def\bstate#1{ {\overline { {| #1\rangle} } } }
\def\vac{ {| 0 \rangle} }
%
\nref\bp{W.~Bardeen and R.~Pearson, Phys.~Rev.~D14 (1976)547.}
\nref\bpr{W.~Bardeen, R.~Pearson, and  E.~Rabinovici,
Phys.~Rev.~D21 (1980)1037.}
\nref\bpa{S.~Brodsky and H.C.~Pauli, Phys.~Rev.~D32 (1985)1993.}
\nref\wiegmann{P.~Wiegmann, Phys.~Lett.~141B (1984)217.}
\nref\bd{M.~Bander and W.~Bardeen, Phys.~Rev.~D14 (1976)2117.}
\nref\phw{R.~J.~Perry, A.~Harindranath, and K.~Wilson,
Phys.~Rev.~Lett. 65 (1990)2959.}
\nref\kss{J.~Kogut, D.K.~Sinclair, and L.~Susskind, Nucl.~Phys.~B247
(1976)199.}
\nref\wz{J.~Wess and B.~Zumino, Phys.~Lett.~37B (1971)95. }
\nref\wzw{E.~Witten, Commun.~Math.~Phys.~92 (1984)455. }
\nref\griffin{P.~Griffin, Fermilab-Pub-91/92-T (1991).}
\nref\schwinger{J.~Schwinger, Phys.~Rev.~128 (1962)2425.}
\nref\bergknoff{H. Bergknoff, Nucl.~Phys.~B122 (1977)215.}
\nref\df{R.~Dashen and Y.~Frishman, Phys.~Rev.~D11 (1975)2781.}
\nref\kz{V.~Knizhnik and A.B.~Zamolodchikov, Nucl.~Phys.~B247
(1984)83.}
\nref\kpsy{D.~Karabali, Q.~Park, H.~Schnitzer, and
Z.~Yang, Phys.~Lett.~216B(1989)307.}
\nref\mccartor{G.~McCartor, Z.~Phys.~C41 (1988)271.}
\nref\go{P.~Goddard and D.~Olive, Int.~J.~Mod.~Phys.~A1 (1986)303.}
\nref\gw{D.~Gepner and E.~Witten, Nucl.~Phys.~B278 (1986)493.}
\nref\hbp{K.~Hornbostel, S.~Brodsky, and H.C.~Pauli, Phys.~Rev.~D41
(1990)3814.}
\nref\blp{W.~Buchm\"uller, S.~Love, and R.~Peccei,
Phys.~Lett.~108B (1982)426.}
\nref\thooft{G.~'tHooft, Nucl.~Phys.~B75 (1974)461.}
\nref\bowcock{P. Bowcock, Nucl.~Phys.~B316 (1989)80.}
\nref\epb{T.~Eller, H.C.~Pauli, and S.~Brodsky, Phys.~Rev.~D35
(1987)1493.}

\preprint{FERMILAB-PUB-91/$197$-T}
\date{August 1991}
\title {Solving $3+1$ QCD on the Transverse Lattice}
       {Using $1+1$ Conformal Field Theory}
\author{Paul A. {Grif}fin}{$^\dagger$}{Internet: pgriffin@fnalf.fnal.gov}
\address{Theory Group, M.S.~106}
\address{Fermi National Accelerator Laboratory}
\address{P.~O.~Box 500, Batavia, IL  60510}
\abstract
A new transverse lattice model of $3+1$ Yang-Mills
theory is constructed by introducing Wess-Zumino terms into the 2-D
unitary non-linear sigma model
action for link fields on a 2-D lattice.  The Wess-Zumino terms
permit one to solve the basic non-linear sigma
model dynamics of each link, for discrete values of the bare QCD
coupling constant, by applying the representation theory of
non-Abelian current (Kac-Moody) algebras.
This construction eliminates the need to approximate the non-linear
sigma model dynamics of each link with a linear sigma model theory, as
in previous transverse lattice formulations.  The non-perturbative
behavior of the non-linear sigma model is preserved by this construction.
While the new model is in principle solvable by a
combination of conformal field theory, discrete light-cone, and
lattice gauge theory techniques, it is more realistically
suited for study with a Tamm-Dancoff truncation of excited states.
In this context, it may serve as a useful framework
for the study of non-perturbative phenomena in QCD via analytic
techniques.

\finishtitlepage

\newsec{Introduction}

The transverse lattice approach to $3+1$ Yang-Mills theory (QCD)
originally developed by Bardeen and Pearson\bp\ over ten years ago,
incorporates conceptual and computational advantages that are found separately
in other formulations.  Like the 4-D Euclidean lattice formulation, the
physical degrees of freedom are link variables of a discrete lattice which are
interpreted as phase factors $e^{i\int A}$.  The transverse
lattice models incorporate the non-perturbative dynamics of QCD
and are well suited for studying the bound state spectrum\bpr.
However in the transverse lattice construction, the lattice is only
two-dimensional.  Local 2-D continuum gauge fields are also present
to gauge the symmetries at each site.
The local gauge invariance is then used to eliminate, via gauge fixing, the 2-D
gauge fields in favor of a non-local Coulomb interaction for the link fields.
This is accomplished in a light-cone gauge
$A_{-}=0$ and with light-cone quantization so that $A_{+}$ can be
eliminated by using its equation of constraint.  All physical states have
positive light-cone energy $P^+$.  This eliminates two degrees of freedom and
simplifies the classification of
the bound states (see ref.~\bpa\ for further discussion of the advantages of
the light-cone approach).

The basic action for each link on the transverse lattice is the 2-D unitary
$SU(N)$ principal chiral non-linear sigma model.
Although this sigma model is exactly solvable via a Bethe-ansatz
technique\wiegmann, this solution cannot be easily applied in the transverse
lattice context.  The Bardeen Pearson model is a linear sigma model
approximation of the non-linear  model in which the unitarity constraint
of the link fields is relaxed.  The $N \times N$ matrices of the linear
sigma model are constrained by introducing potential terms into
the theory which are designed to drive the system into the
non-linear phase\bd.  In the numerical work of Bardeen, Pearson, and
Rabinovici\bpr,
glueballs are constructed from local two-link and four-link bound states
which are smeared over the 2-D lattice.  This truncation of the Hilbert
space is a non-perturbative light-front Tamm-Dancoff approximation to the
QCD bound state problem\phw.
The numerical results based on this approach were inconclusive.  The
links were weakly coupled via the Coulomb interation, and the spectrum
was qualitatively similar to what would be obtained from a strong coupling
expansion in ordinary lattice gauge theory\kss.

There are a number of changes one could make to their original analysis
that might improve the situation.  This paper will focus on
directly solving the non-linear sigma model dynamics instead of using
the linear sigma model approximation.
By introducing Wess-Zumino\wz\ terms into the sigma
model action, we will describe the non-linear sigma model dynamics in the
basis of operators given by the well-studied and exactly solvable
Wess-Zumino-Witten\wzw\ (WZW) model. The WZW
currents will be the linear variables which describe exactly the dynamics
of the non-linear sigma model. The Wess-Zumino terms in the
action will become irrelevant operators in the continuum limit.

The non-linear aspects of the principle chiral sigma model are
retained in the WZW model.  The unitary link fields (and products of link
fields) appear as
the primary fields of the WZW model, and play a crucial role in defining
the highest weight states of the Hilbert space of the WZW model.
The highest weight states correspond to zero modes of
Wilson loops on the transverse lattice.  This zero mode structure is lacking
in the linear sigma model treatment of the transverse lattice
theory.  (The structure presumably corresponds to the space of soliton
excitations of the linear sigma model fields.)

The advantage
of exactly solvable non-linear sigma model dynamics must be weighed against
the two potential disadvantages of this approach.  First, this WZW model
approach will only work for the discrete values of bare sigma model coupling
constants which
correspond to the non-trivial WZW fixed points.  This will in turn place
a constraint on the QCD coupling constants that this model can obtain in the
continuum limit.  These particular values are not special points
in the context of $3+1$ QCD, but rather these are points where we can
apply our limited knowledge of the 2-D non-linear sigma model to simplify
the local dynamics of the link fields.
Second, the continuum limit may be difficult to obtain
because  the irrelevant terms added to simplify the local link dynamics
may be large for finite lattice spacings.  This issue can only be resolved
by explicit numerical simulation.

Preliminary work on the use of the gauged WZW model to describe
the dynamics of lattice model links was discussed in ref.~\griffin.
A transverse lattice model with one lattice dimension
and two continuum dimensions was studied, and
it was found that assigning the same Wess-Zumino term,
with the same coupling constant, to
each link leads to an order $a$ term in the continuum limit, where $a$ is the
lattice spacing.  This order $a$ term generates the $2+1$ pure Chern-Simons
action in the continuum limit,
and its dynamics was discussed in some detail.  The states of the
Chern-Simons model correspond to zero modes of Wilson Loops; similar
states will generate the vacuum sectors in the $3+1$ QCD model.
An important lesson from this work is that one cannot simply
assign the same Wess-Zumino term to each link in the
$3+1$ QCD case, because the leading terms in the continuum limit must
go as order $a^2$ in this case,
and not order $a$ as in the $2+1$ Chern-Simons theory.  This means that the
Wess-Zumino terms must be staggered from site to site, with coupling constants
$\pm k$.  The study of how to correctly stagger the Wess-Zumino terms is the
major topic of this paper.

Section 2 is review of the basic transverse lattice construction of QCD
based on the (unitary) non-linear sigma model.  The
degrees of freedom on the lattice are introduced,
and the ``naive" continuum limit is taken by performing a Bloch wave
expansion of the link fields.
In section 3, we begin the analysis of adding
Wess-Zumino terms to the action.  It is found that the
structure of the staggered Wess-Zumino terms which generate the Coulomb
potential that in turn correctly drives the system to the desired continuum
limit violates local gauge invariance by generating non-Abelian anomaly terms.
In section 4 this difficulty is resolved by
defining a new model which has a different structure of local gauge invariance,
but has the correct continuum limit.
In the new model, pairs
of nearest neighbor links are associated with a single local 2-D gauge
symmetry, and the anomalies from the Wess-Zumino terms for each pair of sites
cancel.  The local 2-D gauge symmetry is reduced by a factor of two
from the transverse lattice construction with no Wess-Zumino terms.  The
remaining local gauge symmetry for each pair of links in the {\it bilocal}
transverse lattice model is then properly
gauge fixed in light-cone gauge.  In section 5,
the current algebraic solution of the WZW model is reviewed and
the quantum theory of the new model is discussed.  Particular
emphasis is placed on the highest weight states of the current algebras,
which generate the space of Wilson loop zero modes on the lattice.  Aspects of
future bound-state calculations and other applications of this construction are
discussed in section 6.
\vfill
\newsec{The Basic Transverse Lattice Construction}

\topfigure {pub91197fig1.eps} {1}
{\ The degrees of freedom associated with each site $\xperp$ are the
2-D gauge fields $A_{\pm} (\xperp )$, and the link fields $U_x (\xperp )$
and $U_y (\xperp )$, defined on the links as in fig.~1.}

In this section, the non-linear sigma model-based formulation of
the transverse lattice construction of QCD is reviewed, and
the process of taking the naive continuum limit is studied.

Consider the matrix-valued chiral fields
$U_\alpha (\xperp; \xplus , \xminus )$, where $\alpha=1,2$,
which belong to the fundamental representation of $SU(N)$.  These fields
lie on the links $[\xperp , \xperp + \bfalpha ]$
of a discrete square lattice
of points $\xperp= a(n_x , n_y )$, with lattice spacing $a$ and basis
vectors $\bfalpha = (a,0)$ or $(0,a)$. The link
fields are continuous functions of the light-cone coordinates
$x^{\pm}=(x^0 \pm x^1 )/\sqrt{2}$, so that the two-dimensional lattice
describes a partially discretized $3+1$ dimensional Minkowski space field
theory\foot{The indexes $\alpha, \beta, \ldots$ denote
transverse coordinates $x,y$, and
$\mu,\nu, \ldots$ denote longitudinal coordinates $x^{\pm}$ .}
(see fig.~1).

The links fields are defined to transform on the left and right
under independent local 2-D gauge transformations associated with the sites
that the links connect,
\eqn\local{
\delta_G U_\alpha =\Lambda_\xperp (x^\mu ) U_\alpha
- U_\alpha \Lambda_{\xperp+ \bfalpha} (x^\mu ) \ .
}
To construct a gauge invariant action, introduce
$SU(N)_{\xperp}$ gauge fields $A_{\pm}(\xperp ) =i A^a_{\pm}(\xperp
)T^a$, where the group
generators $T^a$ satisfy  $[ T^a , T^b ]=
if^{abc} T^c$ and $\Tr\ T^a T^b = \half \delta^{ab}$.
The infinitesimal transformation law for the gauge fields is
\eqn\gaugetran{
\delta_G A_{\pm}(\xperp ) = \del_{\pm} \Lambda_{\xperp}
+ [\Lambda_{\xperp}, A_{\pm}(\xperp ) ]\ ,
}
and the covariant derivative is
\eqn\cdel{
D_{\mu} U_\alpha ( \xperp ) = \del_\mu U_\alpha - A_\mu (\xperp ) U_\alpha
+ U_\alpha A_\mu (\xperp + \bfalpha ) \ .
}
The transverse lattice action is given by\bp
\eqn\tlaction{ \en{
I_{\rm TL}=\sum_{\xperp} \Tr\ & \int d^2 x\Biggl\{ {a^2\over 2g_1^2} F^{\mu\nu}
F_{\mu\nu} + {1\over g^2}\sum_{\alpha}  D_\mu U_\alpha D^\mu
U_\alpha^\dagger\cr
& + {1\over g_2^2 a^2} \sum_{\alpha\neq\beta} \biggl[
   U_\alpha (\xperp ) U_\beta (\xperp + \bfalpha )
   U^{\dagger}_\alpha (\xperp + \bfbeta )
   U^{\dagger}_\beta (\xperp  )- 1\biggr] \Biggr\}   . \cr }
}

As the lattice spacing $a$ is taken to zero,
the interaction terms will select smooth configurations as the
dominant contributions to the quantum path integral;
both the interactions mediated by the local 2-D gauge fields and
the plaquette interactions will
generate large potentials, unless the
link configurations are smooth.  For the plaquette term this is obvious;
for the gauge interactions, this is clear only after
studying the Coulomb potential obtained by gauge fixing in light-cone gauge
$A_- = 0$, in the context of light-cone quantization\bpr.  We will discuss
this process further in the next sections for the new transverse lattice model.
Inserting the Bloch-wave expansion
\eqn\bloch{
U_{\alpha}= \exp \[ -a A_{\alpha} (\xperp + \half\bfalpha ) \]\ ,
}
and keeping only the lowest order contributions, one obtains from
the gauged sigma model kinetic term
\eqn\lowsigma{
I_K = {2 a^2\over g^2} \sum_{\xperp,\alpha} \int d^2 x F_{\mu\alpha}
F^{\mu\alpha}  + \cO (a^4)\ ,
}
and from the the plaquette term
\eqn\lowplaq{
I_P = {a^2\over g_2^2} \sum_{\xperp,\alpha,\beta}\int d^2 x
(F_{\alpha\beta})^2
+ \cO (a^4) \ .
}

In deriving eqn.~\lowsigma, the fields $A_{\pm}(\xperp ) $ were also
assumed to be slowly varying on the lattice.
Combining these three terms and tuning the coupling constants to
$g_1 = g_2 = g$ yields the continuum 4-D QCD action.
For the quantum theory on the lattice,
the Lorentz covariant critical point for each lattice
spacing $a$ is determined by examining specific properties of the
states, such
as the mass spectrum $3+1$ Lorentz multiplets and the covariant
dispersion relations\bpr.

This is the non-linear sigma model (NLSM)-based transverse lattice model of
QCD.
In an ideal world, there would be an exact solution to
the primary chiral sigma model, which could be used as a kernel
to solve the entire model perturbatively in the interactions.
The idea would be to use the states that are diagonalize with respect to the
NLSM Hamiltonian as a basis for construction of the singlet bound states
of the full theory.  This is the approach pioneered by Schwinger in his
solution to $1+1$ massless QED\schwinger.  There, the full Hamiltonian
was diagonalized in the basis of free fermions.
While there has been some progress in understanding
the quantum NLSM\wiegmann, the progress is still insufficient to generate
a Schwinger-type solution to the problem at hand.  A ``Fock space"
of operators with simple commutation relations is required.

\newsec{Transverse Lattice with Wess-Zumino Terms}

Now consider the NLSM with Wess-Zumino term\foot{
Normalization of the kinetic term differs from ref.~\wzw\ because of a
different definition of the trace.  See ref.~\go\ for a thorough discussion
of such normalization issues.}\wzw,
\eqn\action{
I_{WZW} ={1\over \lambda^2}\int d^2 x \Tr\
\del_\mu U \del^\mu U^{\dagger} + k \Gamma\ ,
}
where $U$ is a unitary matrix.
The non-local Wess-Zumino term $\Gamma$ is well-defined only up to $\Gamma
\rightarrow \Gamma +2\pi$, and therefore
the coupling constant $k$ is an
integer.
The model is exactly solvable for the restricted critical values of
the NLSM coupling constant
\eqn\critical{
\lambda^2 = {4\pi \over | k| } \ .
}
For these values there exist a complete basis of conserved vector and
axial-vector currents from which a Fock space representation of the
quantum theory can be constructed.  For positive $k$, the conserved currents
are
\eqn\poscurrents{
J_{-}( x^- ) = (\del_{-}
U  ) U^{\dagger}\ ,\ \ \ J_{+}( x^+ )=
 U^{\dagger} (\del_{+} U ) \ ,
}
and for negative k,
\eqn\negcurrents{
\tJ_{+}(x^+ ) = (\del_{+} U
) U^{\dagger}\ , \ \ \ \tJ_{-}(x^- ) =
U^{\dagger} (\del_{-} U)\ .
}
An elegant current algebraic solution for the quantum theory was given by
Dashen
and Frishman\df\ for $k=1$, and was later generalized for arbitrary $k$ by
Knizhnik and Zamolodchikov\kz.  This solution will be discussed further in
section 5.

To apply the WZW NLSM technology to the transverse lattice formulation of QCD,
we associate a WZW field $U$ with each link $U_\alpha (\xperp )$\foot{The
analysis here generalizes the $2+1$ dimensional transverse lattice
construction given in ref.~\griffin.}.
For each link $U_\alpha (\xperp)$, the currents $J_{\pm,\alpha}$ and
$\tJ_{\pm,\alpha}$ are defined via eqns.~\poscurrents\ and \negcurrents.
The gauge variation of the $J$ currents is given by
\eqn\cvar{
\en{ \delta_G J_{+,\alpha} =&
- \del_+ \Lambda_{\xperp + \bfalpha} + [\Lambda_{\xperp+ \bfalpha},
J_{+,\alpha}] + U^{\dagger}_\alpha \del_+
\Lambda_\xperp  U_\alpha \ , \cr
\delta_G J_{-,\alpha} =&
\del_- \Lambda_\xperp + [\Lambda_\xperp ,
J_{-,\alpha}] - U_\alpha \del_-
\Lambda_{\xperp + \bfalpha} U^{\dagger}_\alpha \  ,\cr}
}
where the currents $J_\pm$ are specified at $\xperp$.  The variations of
the $\tJ$ currents are similar.

The action is invariant under the global symmetry $U\rightarrow A U
B$, where $A,B$ are constant unitary matrices.
The Wess-Zumino term must be gauged so it can be added to
a transverse lattice action.  (The kinetic term of the WZW model
is easily gauged by introducing the covariant derivative as in the
previous section.)  The gauge variation of the Wess-Zumino term is local\wzw\
and can be written in terms of the currents,
\eqn\wzvar{
\en{
\delta_G \Gamma ( U_\alpha (\xperp ) ) = {1\over 2\pi} \int d^2 x \Tr & \biggl[
\Lambda_\xperp \bigl( \del_+ J_{-,\alpha} - \del_-\tJ_{+,\alpha} \bigr) \cr
&- \Lambda_{\xperp+\bfalpha} \bigl( \del_- J_{+,\alpha} -
\del_+\tJ_{-,\alpha} \bigr) \biggr] \ . \cr}
}
It is straightforward to construct the ``gauged" Wess-Zumino term for each
link,
\eqn\gaugewz{ \en{
\tGamma (  U_\alpha (\xperp ) ) =\Gamma & + {1\over 2\pi} \int d^2 x \Tr\
\Biggl\{ \biggl[  A_+ (\xperp ) J_{-,\alpha} - A_- (\xperp +\bfalpha)
J_{+,\alpha}\cr
& + A_+ (\xperp) U_\alpha A_- (\xperp + \bfalpha ) U^{\dagger}_\alpha  \biggr]
-\biggl[  A_- (\xperp) \tJ_{+,\alpha} \cr
& - A_+ (\xperp +\bfalpha) \tJ_{-,\alpha}
+ A_- (\xperp) U_\alpha A_+ (\xperp +\bfalpha ) U^{\dagger}_\alpha
\biggr] \Biggr\} \
.\cr}
}
This term actually is not  gauge invariant, but instead transforms as
\eqn\anomaly{
\delta_G \tGamma (U_\alpha( \xperp)) ={1\over 2\pi} \int d^2 x\Tr
\bigl[ \Lambda_\xperp \epsilon^{\mu\nu}\del_\mu A_\nu (\xperp )
-\Lambda_{\xperp+\bfalpha}\epsilon^{\mu\nu}\del_\mu
 A_\nu (\xperp +\bfalpha )\bigr] \ ,
}
where $\epsilon^{+-}=1$.  This lack of gauge invariance has the
same form as the non-Abelian anomaly in two dimensions, and
is realized at the classical level.  Note that the vector subgroup of
$SU(N)_{\rm left} \otimes SU(N)_{\rm right}$ for each link is anomaly
free\kpsy.
Also, the global gauge symmetries for each site ($x^\pm$ independent gauge
transformations) are unbroken.

Since more than one link is coupled to each site,
the anomaly can cancel between the links.  This is the mechanism
that was introduced in ref.~\griffin\ to cancel the anomalies at each site
of a $2+1$ dimensional transverse lattice model.  For the case at hand,
each link on the 2-D lattice is assigned a Wess-Zumino coupling constant
$k_{\xperp,\alpha}$.  The full action, including the Wess-Zumino term, is given
by
\eqn\tlwzaction{
{\tilde I}_{TL} = I_{\rm TL} + \sum_{\xperp,\alpha} k_{\xperp,\alpha}
\tGamma ( U_{\alpha} (\xperp ) ) \  .
}
and it is anomaly  free for each site $\xperp$ if
\eqn\freedom{
\sum_\alpha \  k_{\xperp,\alpha} -
k_{\xperp-\bfalpha,\alpha} =0\ .
}

\topfigure {pub91197fig2.eps} {2} {\  Up to an overall change of sign, the
figures 2(a)-(c) denote the
anomaly-free vertices allowed for the action ${\tilde I}_{\rm TL}$.  The signs
correspond to Wess-Zumino coupling $\pm k$, where $k$ is a positive integer.}

In the remainder of this paper, we will assign either $+k$ or $-k$ Wess-Zumino
coupling to each link, where $k$ is positive, and the NLSM coupling
constant will be fixed to the critical point $g^2 = 4\pi/ k$.

There are  three pairs of
anomaly free vertices (i.e.,~six total) that can be constructed for each site.
The members of each pair are related by an overall flip of signs,
and representatives of each pair are given in fig.~2.  Each link
is labeled by $\pm$ signs which denote the Wess-Zumino coupling.

The simplest configuration to consider is a lattice of all ``+" links, so that
all vertices are all of the type 2(a).  Unfortunately, this does not work
because
the continuum limit of the gauged Wess-Zumino terms are order $a$,
\eqn\badlimit{
 \sum_{\xperp,\alpha} \tGamma ( U_{\alpha} (\xperp ) ) =
 \sum_{\xperp,\alpha} \Tr\
 \biggl\{   {a\over 2\pi}\bigl[ 2F_{-+}
A_\alpha + A_{+} \del_\alpha A_{-}- A_{-} \del_\alpha A_{+}\bigr]
+ \cO (a^3) \biggr\}
}
This expression, for a one-dimensional lattice, is the pure Chern-Simons
term in $2+1$ dimensions.  It was discussed in some detail in ref.~\griffin.
The leading $\cO (a)$ part comes from the terms in the action that
were added to gauge the Wess-Zumino term $\Gamma$.  The Wess-Zumino term
itself contributes only to order $a^3$, as can be seen by expanding the
variation $\delta \Gamma$ with respect to the link fields $U$ given in
ref.~\wzw.

A possible resolution is to stagger the gauged Wess-Zumino terms from
site to site with alternating signs.  Clearly, there are a number of ways to
stagger these terms.  The ``correct" ways will be those which lead to the right
continuum limit.  In particular, we argued in section 2 that the Coulomb and
plaquette interactions between links would drive the system to a smooth
continuum limit.
With the necessity of staggering,  this is no longer as obvious for
the Coulomb interactions, since the
gauge couplings to each link are not the same on a staggered lattice.
Recall that the Coulomb interactions are mediated by the longitudinal
gauge fields $A_{\pm}$.  These fields can be eliminated in the light-cone
gauge $A_- = 0$, at the expense of generating a non-local Coulomb
potential.  We need to study the form of this potential on a staggered lattice.

In the gauge $A_- = 0$, the part of the path integral which depends
upon the gauge field $A_+$ is
\eqn\apartition{
Z_{\rm GF} =  \prod_{\xperp} [\det \del_- ]_{\xperp} \int [dA_+ (\xperp )]
e^{-i\int d^2 x \big\{ a^2 /2 g_1^2 \ (\del_- A_+ (\xperp) )^2 +A_+ (\xperp )
\cJ_- (\xperp) \big\} } \ ,
}
where $\det \del_- $ is the Fadeev-Popov determinant for each site, and the
currents $\cJ_-  (\xperp)$ are given by reading off the couplings in
eqn.~\tlaction.
The form of the current simplifies dramatically at the WZW critical
points $\lambda^2 = 4\pi / k$.
For these cases,
\eqn\negcur{
\cJ_- (\xperp ) = {k\over 2\pi} \sum_{\alpha^\pm} \bigl\{
J_{-,\alpha^+} (\xperp ) -\tJ_{-,\alpha^-} (\xperp - \bfalpha )
\bigr\} \ ,
}
where the sum over $\alpha^+$ ($\alpha^-$) is over links with
$+k$ ($-k$) Wess-Zumino coupling.
In the context of the WZW model, the currents $\cJ_-$ depend only
upon $x^-$.
The Coulomb interaction is obtained by completing
the square in $A_+$.  After completing the square, the integral over $A_+$
cancels the Fadeev-Popov determinant
and the path integral \apartition\ becomes
\eqn\bpartition{
Z_{\rm GF} = \prod_{\xperp} e^{i\int d^2 x g_1^2 / 2 a^2 \big(
{1\over \del_- } \cJ_- (\xperp ) \big)^2 } \ ,
}
where
\eqn\nonlocal{
{1\over \del_- } \cJ_- (\xperp;x^- ) =\half \del_- \int d y^- |y^- -
x^- | \cJ_-  (\xperp;y^- ) + f_\xperp ( x^+ )\ .
}
The importance of keeping the integration
constant $f_\xperp ( x^+ )$ in the context of the massless Thirring model
was recently discussed in ref.~\mccartor.  In our context,
it is easy to calculate by gauge fixing with the condition $A_+ =0$, thereby
introducing the currents
\eqn\poscur{
\cJ_+ (\xperp ) = {k\over 2\pi} \sum_{\alpha^\pm} \bigl\{
\tJ_{-,\alpha^-} (\xperp ) -J_{-,\alpha^+} (\xperp - \bfalpha )
\bigr\} \ .
}
These currents in the WZW model depend only on $x^+$.  After completing the
square in this case, the path integral \apartition\ is
\eqn\bbpartition{
Z_{\rm GF} = \prod_{\xperp} e^{i\int d^2 x g_1^2 / 2 a^2 \big(
{1\over \del_+ } \cJ_+ (\xperp ) \big)^2 } \ ,
}
where
\eqn\nonlocalb{
{1\over \del_+ } \cJ_+ (\xperp;x^- ) =\half \del_+ \int d y^+ |y^+ -
x^+ | \cJ_+  (\xperp;y^+ ) + {\bar f}_\xperp ( x^- )\ .
}
Equating the two results for the same gauge-fixed path integral yields
\eqn\fequal{
f_{\xperp} (x^+ ) = \half \del_+ \int d y^+ |y^+ - x^+ | \cJ_+
(\xperp;y^+ ) \ ,
}
and the path integral $Z_{\rm GF}=e^{iI_{\rm C}}$,
where
\eqn\coulomb{ \en{
I_{\rm C} = {g_1^2\over 4 a^2}\sum_{\xperp} \int d^2 x \Tr\ \biggl\{
& \int dy^- \cJ_- (x^- ) |x^- - y^-| \cJ_- (y^- ) \cr
+ & \int dy^+ \cJ_+ (x^+ ) |x^+ - y^+| \cJ_+ (y^+ ) \biggr\} \ .\cr }
}
As in the Schwinger model, the Coulomb potential does not mix left- and
right-mover currents.  In the current algebra solution to the
quantum theory discussed in the next section, the Coulomb terms
are treated as potential terms.  The currents $\cJ_{-}$ and $\cJ_+$
then remain functions of $x^-$ or $x^+$ in the WZW model with
Coulomb  interactions.  This is the same situation found
in the massless Schwinger model (see the analysis of
ref.~\bergknoff).

The Coulomb interactions are proportional to $1/a^2$.  As $a\rightarrow 0$,
configurations which minimize the full action should dominate the path
integral.
The question is whether these configurations correspond to the smooth continuum
limit that we desire.
Consider the link $U_{\alpha^+} (\xperp)$.  According to eqns.~\negcur\ and
\poscur, it interacts at $\xperp$ by contributing to the $\cJ_- (\xperp)$
current and interacts at $\xperp + \bfalpha$ by contributing to
$\cJ_+ (\xperp +\bfalpha )$.  Similarly, $U_{\alpha^-} (\xperp)$
interacts at $\xperp$ by contributing to the $\cJ_+ (\xperp)$
current and interacts at $\xperp + \bfalpha$ by contributing to
$\cJ_- (\xperp +\bfalpha )$.
For each of the anomaly-free vertices in figure 2, the links interact
pairwise with each other.  None of the vertices have all four
links contributing to $\cJ_+$ or $\cJ_-$.  Rather, two
links contribute to $\cJ_+$ and two links contribute to $\cJ_-$.
Therefore minimizing the Coulomb interaction as $a\rightarrow 0$
does not necessarily drive the system to the smooth continuum limit that is
required to reproduce continuum QCD.  Further evidence that the vertices
of figure 2 do not generate the correct continuum limit was
obtained by studying the vacuum structure, following the analysis
discussed in section 5 for the correct result.

\newsec{Bilocal Gauge Invariance, and a New Transverse Lattice
Action}

\topfigure {pub91197fig3.eps} {3}
{\  The two vertex configurations for which all four links couple
symmetrically to each other.  In figure 2(a) all four links contribute to
the $\cJ_+$ current at the vertex, and in figure 2(b) all four contribute
to $\cJ_-$.}

The two vertices that have all four links contributing to either
$\cJ_+$ or $\cJ_-$ are given in figure 3.  While these vertices
have the correct behavior under the Coulomb interactions,
they are both anomalous with respect to the local gauge invariance
at each site. Recovering QCD in the continuum limit is our paramount
consideration, so we will consider breaking some of the local gauge
symmetry.  Specifically, we will gauge only the anomaly free local symmetries.
The anomalous local symmetries will be become global symmetries.
The unbroken local symmetries  will then have to be
gauge fixed, and the  coupling of the links via Coulomb interactions
will need to be re-examined.

The solution to the problems of the previous section will make use
of the fact that the two vertices of figure 3 break gauge
invariance in opposite ways.  To be specific, label
each site $\xperp = a(n_x , n_y )$
with the $Z_2$ quantum number
\eqn\lparity{
P_{\rm L} ( \xperp ) = (-1)^{n_x +n_y}
}
which will be referred to as {\it lattice parity} .  Even (odd) sites have
lattice parity $+1$ ($-1$).
As in the previous section, we consider the transverse lattice
action with Wess-Zumino terms, eqn~ \tlwzaction.  However, now we
consider configurations which violate the gauge invariance constraint
eqn.~\freedom.  The assignment of the Wess-Zumino coupling constant
is given by
\eqn\uset{ \en{
U_\alpha (\xperp) &= U_{\alpha^+} (\xperp ) \ , \xperp \
{\rm even} \ , \cr
U_\alpha (\xperp) &= U_{\alpha^-} (\xperp ) \ , \xperp \
{\rm odd} \ , \cr}
}
where the notation $\alpha^\pm$ corresponds to assigning
Wess-Zumino coupling $\pm k$.  The vertex at odd (even) sites is
the type shown in fig.~3(a) (fig.~3(b) ).  The anomaly at each site
is
\eqn\newanomaly{
\delta_G {\tilde I}_{\rm TL} (\xperp ) = P_{\rm L}(\xperp ) {2k\over\pi}
\int d^2 x \Tr\ \Lambda (\xperp ) \epsilon^{\mu\nu} \del_\mu
A_\nu (\xperp )\ .
}
Define nearest neighbor pairs ($\xperp^+ , \xperp^-$),  which by the
above construction have opposite anomalies, as
\eqn\sitepairs{
( \xperp^+ , \xperp^- ) = ( \xperp , \xperp + (-1)^{n_x} {\hat x} )
, \ \ \ \forall \xperp\ {\rm s.t.}\   P_{\rm L} ( \xperp) = 1 \ ,
}
where ${\hat x}=(a,0)$.  Every site on the square lattice belongs to one pair.
By construction, $\xperp^+$ ($\xperp^-$) is an even (odd) site.

For each of these nearest neighbor pairs, the anomaly breaks one of the
local gauge symmetries, and preserves the other local symmetry and
the two global symmetries.  Gauging the local symmetries for this transverse
lattice model no longer requires one independent vector potential
for each site.  Rather, the gauge fields at $\xperp^-$ sites can be
parametrized
as
\eqn\Abreak{
A_\mu (\xperp^- )=G_{\xperp^-} A_\mu (\xperp^+ )
G_{\xperp^-}^{\dagger}\ , }
where $G_{\xperp^-}$ is a constant ($x^\pm$ independent) unitary matrix
which transforms as
\eqn\gtrans{
\delta_G G_{\xperp^-} = \tLambda_{\xperp^-} G_{\xperp^-}\ ,
}
under gauge transformations.  The field $G_{\xperp^-}$ corresponds to the
unbroken global symmetry at the $\xperp^-$
sites.  So instead of having a 2-D gauge field $A_\mu$ for each site,
we now have the set $(A_\mu , G_{\xperp^-})$ for each pair of sites.
The links and vector fields transform as given by eqns.~\local\ and \gaugetran\
as long as the the infinitesimal variations at the $x^-$ sites satisfy the
constraint
\eqn\gaugeconst{
\Lambda_{\xperp^-}(x^\pm ) = \tLambda_{\xperp^-} + G_{\xperp^-}
\Lambda_{\xperp^+}(x^\pm ) G_{\xperp^-}^\dagger \ .
}
With this construction, the gauge variations from the $x^-$ sites cancel the
anomaly from the $x^+$ sites.  The path integral measure is redefined as
\eqn\newmeasure{
\prod_\xperp [dA_\mu ] \rightarrow \prod_{\xperp^+} [d A_\mu
(\xperp^+ )] \prod_{\xperp^-} [dG_{\xperp^-} ] \ ,
}
where $[dG ]$ is the left invariant Haar measure for $G$.
The full action is given by eqn.~\tlwzaction, with the
Wess-Zumino coupling constants given by eqn.~\uset, and the
field identification \Abreak.

It is important that the
local gauge symmetry at each site remain unbroken.  Otherwise, each
pair of links ($U_{\alpha^+} ,U_{\alpha^-}$) would transform the same
way under the remaining local gauge invariance, effectively doubling the
number of link degrees of freedom in the gauge theory, and the theory
would not have QCD as the ``naive" continuum limit.
Each of the remaining local symmetries are associated with two nearest
neighbor sites, paired together as prescribed by equation \sitepairs.
The above construction will be referred to as a {\it bilocal} transverse
lattice model.  Each gauge field $A_\mu ( \xperp^+ ) $ is coupled to
seven links instead of four (see fig.~4).  This difference
is obviously significant at the lattice level.  But again, the point is that
there are many models which have the same continuum behavior which differ
at scales of the lattice spacing.  The bilocal model
has the advantage of being more easily treatable at the lattice level
than the basic transverse lattice model without Wess-Zumino terms.

\topfigure{pub91197fig4.eps} {4} {\  For the bilocal model, each
gauge field $A_\mu (\xperp^+ )$ interacts with seven links.
Figure $4$ shows the case where $\xperp^+$ is to the right of
$\xperp^-$.}

The Coulomb dynamics of the bilocal model is studied by gauge fixing
in the $A_- = 0$ gauge as in
the previous section.  The part of the path integral which depends upon the
gauge field $A_+$ is
\eqn\cpartition{
Z_{\rm GF} = \prod_{\xperp^+} [\det \del_- ]_{\xperp^+} \int [dA_+ (\xperp^+ )]
e^{-i\int d^2 x \Tr\ \big\{ a^2 / g_1^2 (\del_- A_+ (\xperp^+) )^2 +A_+
(\xperp^+ )
\cJ_- (\xperp^+) \big\} } \ .
}
There is an additional factor of two in front of the $( \del_- A_+)^2$ term,
relative to eqn.~\apartition,
from the contribution to the kinetic energy term from the $\xperp^-$ sites.
Note that the $G_{\xperp^-}$s cancel out of this expression.
The current $\cJ_- (\xperp^+ )$ is given by eqn.~\negcur, and all four
links connected to the site $\xperp^+$ contribute to it.
Completing the square yields eqns.~\bpartition\ and \nonlocal, up to the
additional factor of two, and where $\cJ_- (\xperp^- ) = 0$.
The functions $f_{\xperp^+} (x^+ )$ remain to be determined.
Following the previous analysis, we gauge fix in the $A_+ = 0$ gauge
and find
\eqn\dpartition{ \en{
Z_{\rm GF} = & \prod_{\xperp^+} [\det \del_+ ]_{\xperp^+} \int [dA_- (\xperp^+
)]
[d G_{\xperp^-} ]  \cr
& \times e^{-i\int d^2 x \Tr\ \big\{ a^2 / g_1^2 (\del_+ A_- (\xperp^+) )^2 +
A_- (\xperp^+ )
G^{\dagger}_{\xperp^-} \cJ_+ (\xperp^-) G_{\xperp^-}\big\} } \ . \cr}
}
After completing the square and comparing to the $A_- = 0$ case, one finds
\eqn\ffequal{
f_{\xperp^+ } (x^+ ) = \half \del_+ \int d y^+ |y^+ - x^+ |
\cJ_+ (\xperp^- ;y^+ )  \ .
}
The nonlocal Coulomb effective action for the bilocal theory is
\eqn\newcoulomb{ \en{
I_{\rm C} = {g_1^2\over 8 a^2} \int d^2 x \Tr\ \biggl\{
& \sum_{\xperp^+} \int dy^- \cJ_- (x^- ) |x^- - y^-| \cJ_- (y^- ) \cr
+ & \sum_{\xperp^- }\int dy^+ \cJ_+ (x^+ ) |x^+ - y^+| \cJ_+ (y^+ )
\biggr\} \ .\cr }
}
While in the gauge invariant bilocal model action the
$G_{\xperp}$ dependence is required to preserve the global gauge invariance at
$x^-$ sites,  the $G_{\xperp^-}$ dependence cancels in the gauge fixed action
because it is bilinear in the currents.  (The path integral over the
$G_{\xperp^-}$ is finite since the group $SU(N)$ is compact.)
These Coulomb terms have precisely the properties that we desired to obtain the
correct continuum limit.  All links connected to a given site $\xperp$ interact
with each other via \newcoulomb.  While the pairing of sites \sitepairs\
broke a discrete lattice symmetry by differentiating between $x$ and $y$
directions, this symmetry is restored in the Coulomb effective action.
As the lattice spacing $a$ goes to zero,
the Coulomb dynamics drives the system to a smooth continuum limit.

The naive continuum limit is studied, as in the previous sections,
by inserting a Bloch-wave expansion into the gauge invariant action.
Recall that the $A_\alpha$ dependence in \bloch\ was determined by requiring
that it transform as a gauge field,
\eqn\varaa{
\delta_G A_\alpha (\xperp +\half \bfalpha ) = {\Lambda_{\xperp+\bfalpha} -
\Lambda_{\xperp}\over a} +\  [ \Lambda_{\xperp}A_\alpha - A_\alpha
\Lambda_{\xperp+\bfalpha} ] + \cO (a) \ .
}
For the bilocal model it is still possible to meaningfully expand the
$\Lambda$'s as
\eqn\lambdaex{
\Lambda_{\xperp+\bfalpha} -\Lambda_{\xperp}= a \del_\alpha \Lambda_{\xperp}
+ \cO (a^2 ) \ ,
}
since the constraint \gaugeconst\ allows for arbitrary $\tLambda_{\xperp^-}$
transformations at $x^-$ sites, i.e.~the global gauge invariance at each site
is
retained.  $A_\alpha$ transforms as a gauge field as $a\rightarrow 0$ and the
Bloch-wave expansion \bloch\ is valid for the bilocal model.
In the continuum limit, the undesired order $a$
terms \badlimit\ cancel between even and odd sites as $a\rightarrow 0$.  This
is
due to the staggering of the Wess-Zumino terms which is built into the model.
The cancellation of the gauged Wess-Zumino terms between pairs of adjacent even
and odd links is to order $a^3$, because parity prevents the gauged Wess-Zumino
terms from contributing to order $a^2$.  They are therefore irrelevant
operators
in the continuum.

In fact, the order $a$ terms are explicitly cancelled locally in
the gauge fixed lattice action, and there is no need to invoke a cancellation
between sites, as in the gauge invariant analysis.
To see this, expand the currents $J_{\pm}$ or $\tJ_{\pm}$ for each link
order by order in lattice spacing $a$.  The currents $\cJ_+$ and $\cJ_{-}$
given by eqns.~\negcur\ and \poscur\ are order $a^2$, and therefore the Coulomb
interaction \newcoulomb\ for each site is order $a^2$.
The kinetic and plaquette terms for the link fields are also order $a^2$ by the
analysis in section 2.  (The bare ungauged Wess-Zumino terms are order $a^3$
for each link field as discussed in section 3.)

The bilocal transverse lattice model satisfies the primary constraint
that its continuum limit be QCD, at the expense of introducing a somewhat
complicated structure of gauge invariance on the lattice.

\newsec{Quantization of the Bilocal Transverse Lattice Model}

In this section, the quantum theory of the new bilocal transverse lattice model
constructed.
The ``discrete light-cone" approach of ref.~\bpa\ is followed when the
Hilbert space is defined.
The goal of this section is to set up the theory for future computational study
of the bound state problem.

The approach taken is to solve the WZW model for each link and treat the
Coulomb
and plaquette terms as additional potential terms.
The current algebra solution of the WZW model
was first given by Dashen and Frishman\df\ for level $k=1$, and later
generalized to arbitrary level by Knizhnik and Zamolodchikov\kz.
The current algebra is specified by
Sugawara-type algebras for the left and right-moving currents.
For even (odd) links with Wess-Zumino coupling $k$ ($-k$), the currents
are by $J_{\pm}$ ($\tJ_{\pm}$).  For right movers of an even link,
the current commutation relations are
\eqn\sugawara{
[ J_{-}^{a}(x^- ), J_{-}^{b}( y^- )]= i f^{abc}J_{-}^c
( x^- ) -{ i k\over 2\pi} \delta^{ab}\delta^\prime (x^- - y^- ) \ ,
}
where $J_{-}^a$ is\go\
\eqn\jnorm{
\sum_a T^a J^a_- (\xperp ) = { - i k \over \sqrt 2 \pi} J_{-} ( \xperp) \ .
}
The same type of expressions hold for currents $\tJ_-^a$, $J^a_+$ and $\tJ^a_+$
($x^- \rightarrow x^+$ for the $+$ currents).
Left- and right-mover currents commute, as do currents defined for different
links.  These algebras are the equal time commutation relations translated into
light-cone coordinates.  One can specify initial conditions on
the light fronts $x^+ =0$ and $x^- =0$ for the massless currents $J_-$ and
$J_+$ in the WZW model.  However, because of the complicated form of the
plaquette interaction term, initial conditions will be specified on an equal
time surface.

To make the connection between the commutators above and
Kac-Moody algebras, infared cutoffs for
the light-cone coordinates are introduced by hand.
With $x^\pm$  defined on the interval $[-L,L]$, and fields taken to be
periodic, a mode expansion for the currents takes the form
\eqn\modes{\en{
J^a_+ =& {1\over 2L} \sum_n e^{-i\pi n x^+ /L} J^a_n \ ,\ \ \
\tJ^a_+ = {1\over 2L} \sum_n e^{-i\pi n x^+ /L} K^a_n \ ,\cr
J^a_- =& {1\over 2L} \sum_n e^{-i\pi n x^- /L} \bJ^a_n \ ,\ \ \
\tJ^a_- = {1\over 2L} \sum_n e^{-i\pi n x^- /L} \bK^a_n \ ,\cr }
}
where the site dependence of the currents has been suppressed.  The currents
$J^a_n$,
$\bJ^a_n$ are defined for even links, and $K^a_n$,$\bK^a_n$ are defined
for odd links.  The delta function
in the current algebra \sugawara\ is easily defined for period boundary
conditions, and the current algebra is equivalent to the Kac-Moody algebra
\eqn\kacmoody{
[ J^a_m , J^b_n ] = if^{abc}J^c_{m+n} + k m \delta^{ab}\delta_{m,-n}
}
for each of the mutually commuting currents.  The zero modes $J^a_0$
form a subalgebra equivalent to the original $SU(N)$ Lie algebra.

The light-cone singlet vacuum $\vac$, defined for each chiral algebra,
is the unique highest weight state which satisfies
\eqn\vacuum{
J^a_n \vac = 0 \ , n > 0 \ , \ \ J^a_0 \vac = 0 \ .
}
The definition of normal ordering is with respect to this vacuum state,
\eqn\normal{ \en{
:J^a_m J^b_n : = & J^a_m J^b_n \, \ m<0 \ , \cr
	  	=& J^a_n J^b_m \, \ m \ge 0 \ . \cr}
}

For the even link WZW models, the Lorentz generators $P^+ = H+P$ and
$P^- = H-P$ are given by
\eqn\wzwlorentz{\en{
P^+_{\rm WZW}
=& {1\over 2L( 2k + N)}\sum_n :J^a_n J^a_{-n}:\ = {1\over 2L}L_0 \ , \cr
P^-_{\rm WZW}
=& {1\over 2L( 2k + N)}\sum_n :\bJ^a_n \bJ^a_{-n}:\ = {1\over 2L}{\bar L}_0
\ ; \cr}
}
the odd link Lorentz generators are the same with $(J,\bJ)\rightarrow (K,\bK)$.
The modes of the currents are diagonal with respect to the light-cone
momenta, and have non-vanishing commutation relations
$[ L_0 , J^a_{-m} ]= m J^a_{-m}$ and  $[ \bL_0 , \bJ^a_{-m} ]= m \bJ^a_{-m}$.

The space of states for each link is built up by applying a product of raising
operators $ \{ J^a_{-m} \} \{ \bJ^a_{-m} \}$ upon a highest weight (vacuum)
state $\state {\ell_0} \otimes \bstate {\ell_0}$ for the
right-mover and left-mover sectors.  This construction is analogous to the Fock
space basis of the linear sigma model states.
 For the right-mover sector, these states
satisfy
\eqn\hwstate{\en{
J^a_m \state {\ell_0} &= 0 , \ m>0 \ , \cr
E^\alpha_0 \state {\ell_0} &= 0\ , \cr}
}
where $E^{\pm\alpha}_{0}$ and $H^i_0$ are the zero mode currents $J^a_0$ in the
Cartan-Weyl basis of the algebra.  Similar results apply for the left-mover
sector.  The vacuum states are the highest weights of finite dimensional
$SU(N)$ representations $\state {\ell}$ generated by applying zero modes
$E^{-\alpha}_0$. This will be referred to as the zero mode sub-sector of the
full space of states.  The zero modes have non-vanishing zero point momenta
$L_0
\state \ell = \Delta_{\ell} \state {\ell}$ and
$\bL_0 \bstate \ell = \Delta_{\ell} \bstate {\ell}$ where
$\Delta_{\ell} = C_{\ell}/2( N+ k)$, and  $C_{\ell}$ is the quadratic
Casimir of the representation $\ell$.
The representations of the full Kac-Moody algebra for each
highest weight are unitary if the highest weights have Young tableaux with the
number of columns less than or equal to k\gw.

The representations $\state
{\ell}$ are the zero modes of dimensionless primary fields $\phi_\ell (v)$ of
the
chiral algebras.  The primary fields have simple commutation relations with the
currents\df\kz.  For the even links,
\eqn\hwcom{\en{
[ J^{a}_n ,\phi_\ell ( x^+ )]= &e^{i\pi n x^+ / L }\
\phi_\ell (x^+ )\ t^a_{\ell}  \ , \cr
[ \bJ^{a}_n ,\bphi_\ell ( x^- )]= &  e^{i\pi n x^- / L }\  t^a_{\ell}\
\bphi_\ell (x^- )\ , \cr}
}
and for the odd links,
\eqn\hwcomtwo{\en{
[ K^{a}_n ,\phi_\ell ( x^+ )]= &e^{i\pi n x^+ / L }\ t^a_{\ell}\
\phi_\ell (x^+ )  \ , \cr
[ \bK^{a}_n ,\bphi_\ell ( x^- )]= & e^{i\pi n x^- / L }\
\bphi_\ell (x^- )\ t^a_{\ell}\ , \cr}
}
where $t^a_\ell$ is a  generator of $SU(N)$ in the
irreducible representation $\ell$.  Note the ordering difference here between
even and
odd links.  The primary fields are intertwining operators of the
space of states, since they interpolate between vacuum sectors.
Let $\state 1$ denote the fundamental representation of $SU(N)$.  The products
of left-mover and right-mover primary fields $\bphi_{1} \phi_{1}$ for even
links, and
$\phi_1 \bphi_1$ for odd links, are the quantum fields which correspond to the
classical unitary chiral field $U$ in the classical action.  Since the
classical
unitary field transforms on the left and right in the same
representation, we will consider only the diagonal\gw\ products of the
highest weight fields as the vacuum states in the transverse lattice theory.
The reader is referred to refs.~\wzw\kz\gw\ and in particular the review
article
ref.\go\
for further information on the WZW model.

The gauge fixing proceedure of the previous section left the global symmetry
at each site untouched.  The generators of these gauge transformations in the
quantum theory are constructed from the zero modes of the currents
\eqn\globall{\en{
\cJ^a_0 (\xperp^+ ) & = \sum_\alpha \biggl[ \bJ^a_0 (\xperp^+ , \alpha )-
\bK^a_0 (\xperp^+ - \bfalpha , \alpha ) \biggr]\ , \cr
\cJ^a_0 (\xperp^- ) & = \sum_\alpha \biggl[ K^a_0 (\xperp^- , \alpha )-
J^a_0 (\xperp^- - \bfalpha , \alpha ) \biggr]\ . \cr}
}
In the classical theory, all physical states are gauge invariant.  In the
quantum theory, this restriction can be relaxed somewhat, since it is the
gauge invariance of expectation values of states (operators) that is
required\foot{A modern example of this treatment of gauge symmetries
is found in string theory, where conformal invariance is a crucial property of
the quantum theory.  The Virasoro modes $L_n$ generate
conformal transformations, and physical states need to be annihilated by only
the modes with $n\ge 0$.}.
Following the previous treatment of this point for the the 1-D transverse
lattice\griffin, we require
\eqn\physical{ \en{
 \cE^{\alpha}_0 \state {\rm physical} &= 0 \ , \cr
 \cH^i_0 \state {\rm physical} &= 0 \ , \cr}
}
where $\cE^{\pm\alpha}_0, \cH^i_0$ denote $\cJ^a_0$ in the Cartan-Weyl basis.
In the 1-D transverse lattice case, these constraints led to the correct
set of physical states.

\topfigure {pub91197fig5.eps} {5}
{\  The two cases encountered when gluing together the zero modes of
two links to satisfy the global gauge invariance constraints at a site.
Figure 5(a) \ (5(b)) denotes two links of the same (opposite) lattice parity.}

This criterion can first be applied to the subspace
of states obtained by taking products of zero mode states on the lattice.
In the 1-D transverse lattice construction, non-trivial states of this type
were
found.  They were interpreted as zero modes of Wilson loops on the lattice.
Each link is associated with a product of left-mover zero modes
$\state \ell$ and right-mover zero modes $\bstate \ell$.  Even and
odd links at $U_\alpha (\xperp )$ have zero mode structure
\eqn\zerolinks{\en{
& {\bstate \ell}_{\xperp,\alpha} \times {\state \ell}_{\xperp,
\alpha} \ \ , {\rm even \ link}\ , \cr
&{\state \ell}_{\xperp, \alpha} \times {\bstate \ell}_{\xperp,
\alpha} \ \ , {\rm odd \ link}\ .  \cr}
}
Consider gluing together two links with non-trivial zero mode structure at a
site $\xperp$, such that \physical\ is satisfied.  There are two basic
cases to consider, as shown in figure 5.  In figure 5(a) (5(b)), the two
links have the same (opposite) lattice parity.  In figure 5(a), two
right-mover zero modes $\bstate m \times \bstate \ell$ need to be glued
together.
The constraints \physical\ are satisfied if we project onto the singlet
sector of the tensor product: $P^0_{m,\ell} \{ \bstate m \otimes \bstate \ell
\}$.
This is because the constraint involves the sum of currents for each link,
i.e.~the diagonal subalgebra of the link zero mode algebras.  For
the other case denoted in figure 5(b), the constrains involve the difference
between the currents for each link.  This was the situation encountered
in the 1-D transverse lattice analysis.  The solution to the constraints is to
project onto
the highest weights of the same representation: $\state {\ell_0} \times
\state {m_0} \ \delta_{\ell,m}$.  These two rules for gluing together zero
modes
at a site can be used to construct a wide variety of solutions to the
gauge invariance conditions.  For example, a plaquette solution is of the form
\eqn\zerobox{ \en{
&P^0 \biggl\{ {\bstate {\ell^\dagger}}_{\xperp^+ , y} \otimes
{\bstate {\ell} }_{\xperp^+ , x} \biggr\}\  {\state {\ell_0}}_{\xperp^+ , x}
\ {\state {\ell_0}}_{\xperp^+ + {\hat x} , y} \cr
\times  &P^0 \biggl\{ {\bstate {\ell}}_{\xperp^+ +{\hat x} , y} \otimes
{\bstate {\ell^\dagger }}_{\xperp^+ +{\hat y} , x} \biggr\}\ {\state
{\ell^\dagger_0}}_{\xperp^+ +{\hat y} , x}
\ {\state {\ell^\dagger_0}}_{\xperp^+  , y} \ , \cr}
}
where $\ell^\dagger$ is the conjugate representation of $\ell$.
This plaquette state is the zero mode of a Wilson loop with
flux in representation $\ell$, flowing in the counterclockwise orientation.
It has zero point energy $(P^+_{\rm WZW} + P^-_{\rm WZW} )/2 =
2\Delta_{\ell}/L$
and momentum
$(P^+_{\rm WZW} = P^-_{\rm WZW} )/2 = 0$.  This is the part of a Wilson loop
that can never be gauged away.  While in the linear
sigma model transverse lattice theory\bp\bpr\ these type of states are
presumably
soliton excitations, in the new non-linear theory they arise quite naturally
as vacuum sectors.  The space of states for the new transverse lattice
model breaks up into different vacuum sectors of Wilson loop zero modes, and
the intertwining operator, which takes states from one sector to another,
is the plaquette operator discussed below.

The contribution of the Coulomb interactions to the Lorentz generators
is obtained by inserting the mode expansions \modes\ into the effective
action \newcoulomb\ and normal ordering with respect to the vacuum
state \vacuum,
\eqn\coulorentz{\en{
P^+_{\rm C} =& 2L \biggl( {g_1 \over 8\pi a}\biggr)^2 \sum_{ \xperp^- }
\sum_{n=-\infty}^{\infty} {1\over n^2} :\cJ^a_n (\xperp^- )
\cJ^a_n (\xperp^- ): \ , \cr
P^-_{\rm C} =& 2L \biggl( {g_1 \over 8\pi a}\biggr)^2 \sum_{ \xperp^+ }
\sum_{n=-\infty}^{\infty}  {1\over n^2} :\cJ^a_n (\xperp^+ )
\cJ^a_n (\xperp^+ ): \ , \cr}
}
where
\eqn\calmodes{\en{
\cJ^a_n (\xperp^+ ) =& \sum_{\alpha}  K^a_n (\xperp^+ ,\alpha )
- J^a_n (\xperp^+ -\bfalpha ,\alpha )\ , \cr
\cJ^a_n (\xperp^- ) =& \sum_{\alpha}  \bJ^a_n (\xperp^- ,\alpha )
- \bK^a_n (\xperp^- -\bfalpha ,\alpha )\ . \cr}
}
Note that the light-cone momenta $P^\pm_{\rm C}$ are proportional to the
infared
cutoff $L$.
The contribution to the mass operator from the Coulomb potential
$M^2_{\rm C} =P^+_{\rm WZW} P^-_{\rm C} + P^-_{\rm WZW} P^+_{\rm C}$ is
therefore independent of $L$.  Diagonalization of $M^2_{\rm C}$ on a basis
of states defined in the cutoff theory is therefore the exact (cutoff
independent) result in that basis.  The masses are finite only if the
potentially infared divergent $n=0$ coefficients in $P^\pm_{\rm C}$ vanish
\eqn\zerovanish{
:\cJ^a_0 (\xperp ) \cJ^a_0 (\xperp ): =0 \ , \ \ \ \forall \ \xperp \ .
}
This is a statement of charge neutrality for each site.  The
incoming charge must equal the outgoing charge, where the direction
is defined by the arrows associated with each link (See figure
4).  The zero mode states discussed above, and in particular the
state given by eqn.~\zerobox, satisfy this constraint by construction.
In fact, the constraints \zerovanish\ and \physical\ are equivalent.
For a localized state, such as a link-antilink excitation, the constraint
requires that physical states be gauge singlets at each site.

The structure of the Coulomb term for each site $\xperp$ is similar to that
obtained in $1+1$ QCD with massless fermions quantized
following the same approach\hbp.  In that case, it is known that there
is no mass gap\blp\ generated by the Coulomb interactions, because states can
be
constructed from the $U(1)$ fermion number current
which commutes with the non-Abelian currents that make up the Coulomb
potential.
This is not the case for the new transverse lattice model, since this current
does not exist in this case.  Analysis of the
'tHooft equation\thooft\ for link-antilink bound states can make this result
quantitative by determining the bare mass gap for these states.  Preliminary
calculations show that a linear Regge trajectory for the mass spectrum is
obtained in the large $N$ limit.

The plaquette interaction in \tlaction\ explicitly mixes left-mover and
right-mover modes, like a mass term.  In principle, then, one could
eliminate left-movers in terms of right-movers by solving a constraint equation
for each equal light-cone time surface $x^+ = {\rm const}$.  However, in
practice this is very difficult because of the complicated form of the
interaction in terms of the WZW currents.  Therefore, the plaquette term will
be treated as an interaction in the quantum theory with independent
commutation relations for both left- and right-movers.  The relevant
Cauchy surface on which to fix initial conditions is an equal time surface.
The WZW model current algebras are equal time commutation relations written
in light-cone coordinates (see in particular sec.~2 of ref.~\bowcock.).

For fixed time quantization, the link fields are functions of the single
variable $x$, and for $t=0$ it is useful to define the complex variable
$z = e^{i \pi x /\sqrt 2 L}$.  The quantum link fields are products of
left-mover
and right-mover highest weights
\eqn\qlink{ \en{
U ( z ) =& : \bphi_1 ( z^* ) \phi_1 ( z ) :\ ,\ \ \
{\rm even\  link}\ ,\cr
U ( z ) =& : \phi_1 ( z  ) \bphi_1 ( z^* ) :\ ,\ \ \
{\rm odd\  link}\ ,\cr}
}
where $\phi_1$ is the primary field in the fundamental representation of
$SU(N)$.
The conjugate $U^\dagger$ can be similarly defined in terms of the conjugate
primary fields $\phi^\dagger_1$.  The contribution of the plaquette
interaction to the Lorentz group generators is then
\eqn\boxlor{
P^\pm_{\rm P} =   {L\over \sqrt 2 \pi g_2^2 a^2}
\oint dz \sum_{\beta\neq\alpha} \Tr\
 \biggl[ 1 -   U_\alpha (\xperp ) U_\beta (\xperp + \bfalpha )
   U^{-1}_\alpha (\xperp + \bfbeta )
   U^{-1}_\beta (\xperp  ) \biggr]  \ .
}
The trace in eqn.~\boxlor\ denotes a projection onto products of
highest weights for each site such that the gauge invariance
constraints are satisfied.
The plaquette interaction does not contribute to the spatial momentum
$(P^+ - P^- )/2$.

\newsec{Discussion}

Like any other theory which describes non-perturbative behavior
of QCD, the transverse lattice construction outlined above is very
complicated.  The
basic advantage over 4-D lattice simulations is that a continuum analysis is
used to describe local link dynamics.  To take advantage of the continuum
description, one has to find a suitable truncation of the full model
that still contains the desired physics.  For the
calculation of glueball masses in in the Bardeen Pearson model\bpr, the basic
degrees of freedom were truncated to link-antilink and four-link bound states.
These states mix under the plaquette interaction, which also provides for
transverse motion.
It is essential to complete this basic analysis for the new transverse lattice
model and verify that in this case the link number expansion is a valid one,
i.e.,~that link number violation is strong enough to allow for transverse
motion of the states, yet small enough to validate the link number expansion.

The physical state of the link-antilink and four-link truncated basis is a
Wilson loop smeared over the lattice.  The link-antilink states correspond
to Wilson loops which extend in the longitudinal directions.  This
type of state in the gauge fixed transverse lattice model is the
bilinear $U(x) U^\dagger (y)$ integrated over a wavefunction.  The
link-antilink bound state is {\it not} a bilinear $J^a J^a$ of WZW currents.
Although the currents are the linear degrees of freedom of the WZW model, the
real degrees of freedom of the lattice theory are the link fields.

The spectrum of bound states that can be constructed by the above formalism is
degenerate because the zero mode of the state can be shifted and boosted by
Lorentz transformations.  We want to select a basis of bound states that
does not contain copies of the same state.
Moreover, we want a basis of states for which one of the momentum operators
is manifestly diagonal.  For instance, in the linear sigma model analysis\bpr,
$P^-$ was manifestly diagonal.  This type of analysis in the
bilocal model is complicated, because as discussed above both the left- and
right-mover currents are treated as dynamical.  There is no simple constraint
equation which can be used to solve for one set in terms of the other.

There is a basis of states one can use to study, in the
simplest way, the bound state spectrum.  It is defined by allowing zero-mode
excitations for both left- and right-movers, but truncating all non-zero mode
excitations of the left-movers.  (The parity conjugate basis is equally
simple.)
There is no proof that
all physical states have a representative is this truncated basis.
However, this truncation is similar to the analysis of ref.~\epb, where the
spectrum of the massless Schwinger model was studied.  There,
bound states were constructed from fermions of only one chirality.
In this chiral Schwinger model, there exists only one copy of the free massive
scalar boson, instead of the infinite number of degenerate copies
which exist in the full Schwinger model\mccartor.  Nevertheless, the
single copy has the correct mass.
Again, the idea here is to suggest a starting point for the explicit
calculation
of the mass spectrum.

In the truncated basis, a local link-antilink state is given by
\eqn\linkme{
\state P = \oint dz \oint dy \sum_{k=0}^{P} w(k) \ z^{-k} y^{k-P}\
\Tr :\phi_1 (z) \phi_1^\dagger (y) : \state 0 \otimes \bstate 0 \ .
}
For the link-antilink state, the only possible
right-mover singlet state is the vacuum.  Periodicity in the $z$ and $y$
variables quantizes the momenta $P$ and $k$ to be integers.  The diagonal
contribution to the light-cone momentum $P^+$ is from the WZW model.
Using the conformal field theory commutator $[ L_0 , \phi_1 (z) ] = (z\del_z +
\Delta_1 )\phi_1$, one finds
\eqn\pplus{
P^+_{\rm WZW} \state P = {P + 2\Delta_1 \over 2 L } \state P \ .
}
In the truncated basis, the contribution $P^+_C$ from the Coulomb interactions
always vanishes.  And as discussed in the previous section, the
contribution $P^+_{\rm P}$ from the plaquette interation mixes this state with
four link states.  A physical two link state is a local state\linkme\ smeared
over the entire lattice with a wavefunction specifying the transverse momentum
distribution.

The simplest non-trivial local four link state is obtained by replacing
the right-mover zero modes in equation \zerobox\ with integrals over the
link fields.  There are actually a number of local four link configurations to
consider (see fig.~2 of ref.~\bpr).  The reader is referred to ref.~\bpr\
for a complete description of the method for determining the mass
spectrum of the $3+1$ Lorentz multiplets in the two and four link basis.

The $U(1)$ case avoids all of the complications developed
in sections 3 and 4, and therefore may be a good laboratory to probe the
recovery of 4-D Lorentz invariance from the transverse lattice construction.
The non-linear sigma model action is then the Gaussian model for the
fields $\theta$, where $U=\exp i \theta$.  The plaquette interaction
is a product of normal ordered $U(1)$ vertex operators.  However, it may
be difficult, if not impossible, to approach the continuum limit for the
Abelian
case, since link number violation will be large for a deconfined theory.

Unlike 4-D lattice gauge theory, the transverse lattice construction may
be able to generate structure functions of relativistic bound states,
since the wavefunctions of the states are explicitly calculated when
diagonalizing the mass spectrum.  Only when such a problem, impossible to
work out by conventional techniques, is solved via the transverse lattice
construction, will this new approach be fully accepted as a tool for probing
non-perturbative physics.

The transverse lattice construction connects
2-D physics to more realistic higher dimensional models.  Since string theory
has motivated a great deal of recent progress in 2-D field theory, there are
surely many more connections that can be made, to the benefit of both
mathematically- and phenomenologically-oriented physicists.

\ack I would like to thank S.~Brodsky, L.~Day, O.~Hernandez, K.~Hornbostel,
A.~Kronfeld, J.~Lykken, P.~MacKenzie, S.~Pinsky, and H.C.~Pauli for
numerous useful discussions.  Some of this work was completed at the Aspen
Center for Physics and at the Max-Planck Institut f\"ur Kernphysik in
Heidelberg, Germany.
I am particularly grateful to
W.~Bardeen for many illuminating discussions on transverse lattice physics.

\vfill\eject
\listrefs
\bye